# MEMS Microphones as Ultrasonic Transducers: Nonlinear Electrostatic Actuation and a Parametric Array Prototype

**Xiaoyu Niu, Zihuan Liu, Ehsan Vatankhah, Yuqi Meng, Neal Hall**
Department of Electrical and Computer Engineering, The University of Texas at Austin, Austin, Texas, 78712, USA; xyniu@utexas.edu; zihuanliu@utexas.edu; e.vatankhah@utexas.edu; yuqimeng@utexas.edu; nahall@utexas.edu

## 1. ABSTRACT

This paper investigates commercial-style capacitive MEMS microphone dies as air-coupled ultrasonic transmitters under nonlinear pull-in and snap-back actuation and demonstrates a compact parametric-array prototype. A single die produces large diaphragm displacement and measurable ultrasonic pressure in air. A 28-die array driven at 83 and 93 kHz generates a directional component at the 10 kHz difference frequency. Measurements are compared with analytical radiation theory and finite-element modeling, and the effects of aperture, fill factor, device uniformity, and receiver nonlinearity are discussed.

## 2. BACKGROUND

Capacitive micromachined ultrasound transducers (CMUTs) and piezoelectric micromachined ultrasound transducers (PMUTs) can support air-coupled ultrasound applications including gesture recognition[1], sensing, imaging[2], range finding[3], and haptics[4]. In many applications, a relevant figure-of-merit is achievable transmit pressure. Achieving high transmit pressure requires large diaphragm displacement. PMUTs have proven capable of achieving large displacement and therefore large sound pressure levels (SPLs)[4–6]. Intrinsic stress from piezoelectric film deposition can challenge PMUT manufacturing uniformity. Immersion CMUTs have exploited collapse mode[7–9], deep collapse mode, collapse-snapback mode[10], and other non-collapse nonlinear drive schemes[11–13] to achieve improvements in transmit pressure and to optimize CMUTs for various applications. Applications of such techniques are less explored for air-coupled devices, with a few notable exceptions. Studies[14,15] demonstrate bias-free drive signals in a non-collapsed mode of operation. Towards the development of digital CMOS micro-speakers, electrostatic MEMS transducers were used to generate audio-band pressures in air using a collapse-snapback mode of operation.[16] A commercial MEMS microphone structure was used to demonstrate collapse-snapback mode.[17] The large diaphragm displacement was captured by laser Doppler vibrometer (LDV). Transient ultrasonic pressure waveforms resulting from pull-in and snap-back phenomena have been measured. These studies[18,19] explain collapse-snapback mode vibration in detail, predict diaphragm displacement using a state-space model, and showcase high SPL resulting from large diaphragm displacement. They[17–19] all examine a single MEMS microphone. To the authors' knowledge, this article not only presents intense ultrasound generated by a single CMUT operated in collapse-snapback mode but also provides the first characterization of an array assembled from MEMS microphone dies operated in this mode. Such an array paves the way for other interesting acoustic applications. The conference report and complementary airborne-ultrasound LDV measurements provide additional context for this work.[34,35] Related studies from our group have also examined optical, AlN, lithium-niobate, and multimodal piezoelectric microphone and ultrasonic-transducer platforms.[36–41,43–45]

Parametric array (PA) has been an interesting acoustic phenomenon since it was first demonstrated by Bennett and Blackstock in the 1970s.[20] Loudspeakers were invented more than one hundred years ago. Developments in nonlinear acoustics enable sound to be beamed in a specific direction. Holosonic Research Labs began commercializing parametric-array directional speakers in the early 2000s.[21] Afterwards, other companies entered the market, e.g., Turtle Beach Corporation, Ultrasonic Audio Technologies, Audfly Technology (Suzhou), Rchard Haberkern, etc. However, the PA products are too large and power-consuming for consumer electronics (CE), e.g., cellphone, tablet, and laptop. Wygant et al.[22] built a PA device using vacuum sealed CMUTs. Ahn et al.[23] built one using PZT PMUTs. Both studies demonstrated the feasibility of MEMS-based parametric arrays. However, driving voltage, size, and manufacturing uniformity remain challenges for building a MEMS PA for CE applications. We built a PA device using a CMUT array actuated in collapse-snapback mode. This actuation mode avoids resonance effects, so large acoustic pressure in air is easier to manipulate. Collapse-snapback CMUTs may offer a solution for CE applications.

# 3. NONLINEAR ELECTROSTATIC ACTUATION

Figure 1(a) presents a photograph taken from the top side of a MEMS microphone die. The wire-bonded die is shown at the top of the photograph. The diameter of the circular diaphragm is 1 $mm$. Figure 1(b) presents the cross-sectional schematic following the red dashed line in Figure 1(a). Features of the cross section are common to many commercial MEMS microphones. A relatively thick and perforated backplate serves as a stiff top electrode. The backplate is formed of tensioned silicon nitride with an embedded metal electrode. The backplate contains dimples which are used to prevent the diaphragm from permanently sticking to the backplate upon pull-in by minimizing the contact surface area. The total travel range of the diaphragm in operation is limited to 1.8 $\mu m$, which is the gap distance between the diaphragm and the dimples on the backplate. The thin silicon nitride diaphragm has a metallic surface electrode patterned to mirror that of the top electrode. The diameters of the diaphragm and electrode region are 1 $mm$ and 650 $\mu m$, respectively. In the photograph in Figure 1(a), the electrode embedded in the backplate is visible from above as the relatively bright central region because silicon nitride is semitransparent.

Figure 1(c) presents a classical lumped-parameter model for electrostatic transduction. Unlike Osterberg's model schematic[24], we invert the positions of the diaphragm and backplate to mimic MEMS microphones' realistic structure. The top and bottom electrodes represent the backplate and diaphragm, respectively. Note that the top electrode is fixed in the backplate and immovable. Bottom electrode is embedded in the diaphragm, which is flexible. Equation 1 represents the net force $F$ on diaphragm. Assuming opposite charges on the two electrodes, the diaphragm is attracted upward by electrostatic force; meanwhile, the diaphragm restoring force acts downward.

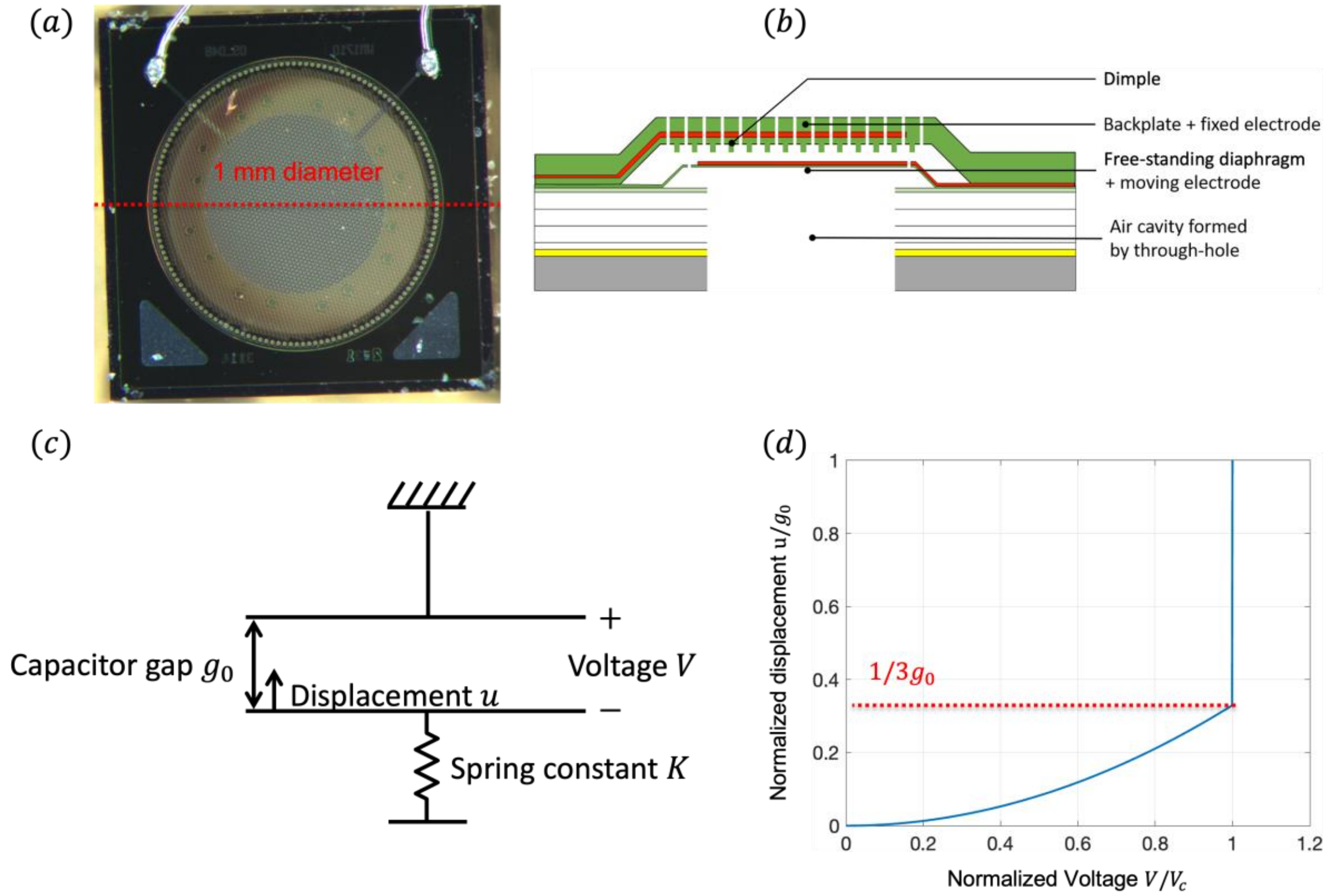

***Figure 1 (a) A photograph taken from the top side of the MEMS microphone die. (b) A cross-sectional schematic illustration of the MEMS microphone die. (c) A schematic of electrostatic transduction. (d) Normalized diaphragm displacement in response to the static normalized voltage.***

$$F(u) = Ku - \frac{\frac{1}{2}(\varepsilon A)}{(g_0 - u)^2}V^2 \tag{1}$$

The net force, defined as positive upward, is equal to the diaphragm restoring force minus the electrostatic force. $u$ is diaphragm displacement. $K$ is spring constant (i.e., diaphragm stiffness). $\varepsilon$ is permittivity between the two electrodes. $A$ is the electrode area. $g_o$ is uncharged gap between the two electrodes. $V$ is the electrical potential between the two electrodes.

When the static voltage across the two electrodes is relatively small, the diaphragm can reach a location with zero net force, i.e., the diaphragm restoring force equals the electrostatic attraction. The system reaches force equilibrium. Note that $\frac{\partial F}{\partial u}$ must be smaller than zero to keep the system stable. On the other hand, as the static voltage increases, the electrodes move closer together. The electrostatic force increases faster than diaphragm stiffness does because electrostatic force is inversely proportional to $(g_0 - u)^2$ but diaphragm stiffness is proportional to $u$. $\frac{\partial F}{\partial u}$ reaches zero when voltage is large enough. At this point, the diaphragm is fully pulled in (collapsed). Similarly, as the voltage decreases, electrostatic force decreases faster than diaphragm stiffness does. The diaphragm returns toward equilibrium. Snap-back occurs. The pull-in and snap-back phenomenon is common to any type of electrostatic device. At the instant when the diaphragm pulls in, both Eqs. (2) and (3) are satisfied simultaneously.

$$F\left(u_p\right) = 0 \tag{2}$$

$$\frac{\partial F}{\partial u_p} = 0 \tag{3}$$

$u_p$ and $V_p$ are the pull-in displacement and voltage, respectively. By solving those two simple equations, two classical values are given in Eqs. (4) and (5). In this case, we do not take intrinsic stress into consideration to simplify the physical interpretation. In practice, in most MEMS devices, intrinsic stress cannot be ignored and may even dominate.

$$u_p = \frac{1}{3} g_0 \tag{4}$$

$$V_p = \sqrt{\frac{8K g_0^3}{27\varepsilon}} \tag{5}$$

Figure 1(d) plots the diaphragm response in response to voltage. Note that diaphragm displacement changes approximately linearly with voltage when voltage is small. In other words, electrostatic transduction is linear when diaphragm displacement is smaller than one third of the total gap distance between two electrodes. Most MEMS devices operate in this region. However, once voltage exceeds the pull-in voltage, diaphragm is pulled into contact with the backplate. Large nonlinearity is induced. Most MEMS devices avoid inducing such nonlinearity. Conversely, we intentionally operate our device in this region to obtain the largest diaphragm displacement.

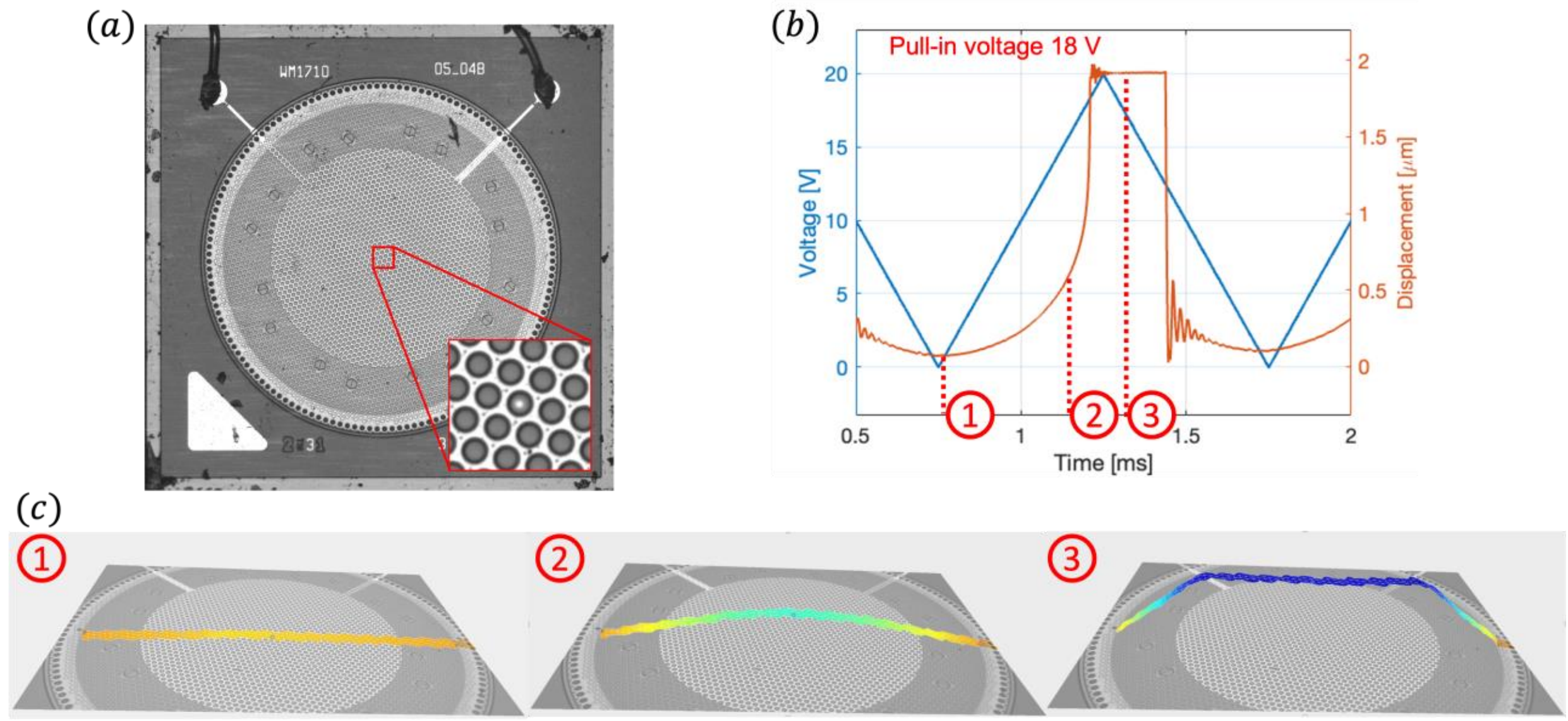

***Figure 2 (a) Center diaphragm displacement in response to a triangular-wave excitation. (b) LDV setup for center-displacement measurement. (c) diaphragm band laser scan at corresponding moments shown in (a).***

To observe the pull-in and snap-back, the center-point displacement was measured using a Polytec MSA-600 laser Doppler vibrometer (LDV). Figure 2(b) shows the LDV measurement point. The approximately 1 $\mu m$ diameter laser spot of the MSA-600 passes through a backplate perforation hole to reflect directly from the diaphragm. The diaphragm of the particular MEMS microphone under study experiences pull-in instability at approximately 18 $V$. This is observed in Figure 2(a), which presents the center diaphragm displacement in response to a triangular-wave input at 1 kHz and ranging in signal value from 0 to 20 $V$. Figure 2(c) presents a line scan of a pull-in process. The three profiles in Fig. 2(c) correspond to the instants labeled in Fig. 2(a). To measure motion along a line, we recorded the response at each point under the same actuation and reconstructed the displacement profiles at the three instants. At the third instant in Fig. 2(c), the diaphragm is flattened, indicating that it has fully collapsed.

## 4. LARGE ACOUSTIC PRESSURE

A goal of this work is to demonstrate large-amplitude diaphragm-displacement waveforms and resultant acoustic pressure waveforms in air. For the first goal, a tone burst signal was applied as shown by the blue trace in Figure 3(a). The signal is a 4-cycle square wave burst with a frequency equal to 4 kHz and with the signal value ranging from 0 to +20 $V$. The resulting displacement measurement is shown by the red trace in Figure 3(a). The diaphragm traverses the full 1.8 $\mu m$ gap upon pull-in and snap-back. The multi-degree-of-freedom state-space model[18,19] predicts the diaphragm displacement shown by the yellow trace in Figure 3(a). Figure 3(b) shows a burst with an increased frequency equal to 36 kHz. Here again, the diaphragm traverses the full 1.8 $\mu m$ gap. In this case, the ringdown of the diaphragm during the release cycle is interrupted by each subsequent pull-in cycle. Figure 3(c) presents the displacement waveform resulting from a 4-cycle burst applied at 96.8 kHz, which is well beyond the fundamental diaphragm resonance frequency of 45.35 $kHz$. The period of the excitation in this case is shorter than the transient pull-in time of the diaphragm, and the diaphragm therefore does not have time to traverse the full 1.8 $\mu m$ gap.[17] Recent high-velocity LDV measurements further illustrate nonlinear MEMS operation at large vibration amplitudes.[42]

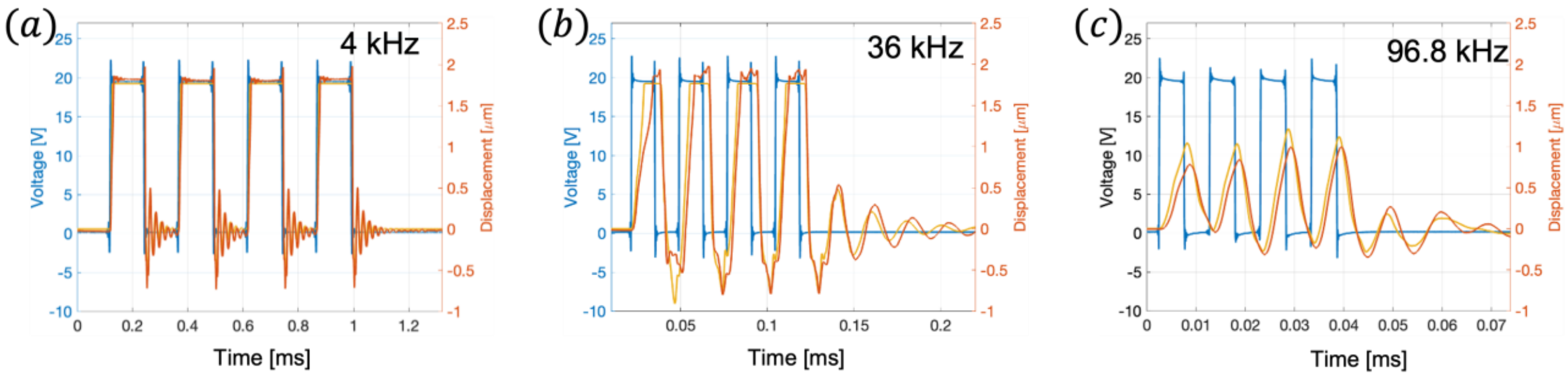

***Figure 3. Tone-burst actuation and diaphragm displacement at 4 kHz (a), 36 kHz (b), and 96.8 kHz (c). The blue trace is voltage actuation oscillating from 0 to 20 V. The red trace is diaphragm displacement measurement. The yellow trace is diaphragm displacement modelling.***

To estimate the resulting large acoustic pressure, Eq. (6) gives the surface acoustic pressure at the diaphragm center $p_c$. $u_c$ is displacement at the center of the diaphragm. $z_0$ is characteristic impedance of air, equal to $\rho_0 c_0$.

$$p_c = (2\pi f u_c) z_0 \tag{6}$$

At $4\ kHz$, $p_c = (2\pi \times 4\ kHz \times 1\ \mu m) \times 1.2 \frac{kg}{m^3} \times 344 \frac{m}{s}$ and $SPL = 20lg\left(\frac{\frac{p_c}{\sqrt{2}}}{p_{ref}}\right) = 111\ dB$. Reference pressure $p_{ref}$ is $20\ \mu Pa$. Similarly, $SPL$ is equal to 130 $dB$ at 36 $kHz$; $SPL$ is equal to 137 $dB$ at 96.8 $kHz$. Note that we distinguish displacement at center of diaphragm $u_c$ and area-averaged displacement of diaphragm $u_d$. To study a single MEMS microphone, this distinction has little effect when emphasizing the large displacement of a single MEMS microphone. Yet, $u_c$ or $u_d$ makes a substantial difference for calculating acoustic pressure of an array. The two quantities must therefore be clearly distinguished. Conventionally, an effective coefficient is used $\alpha$ to describe the conversion, i.e., $u_d = \alpha u_c$.

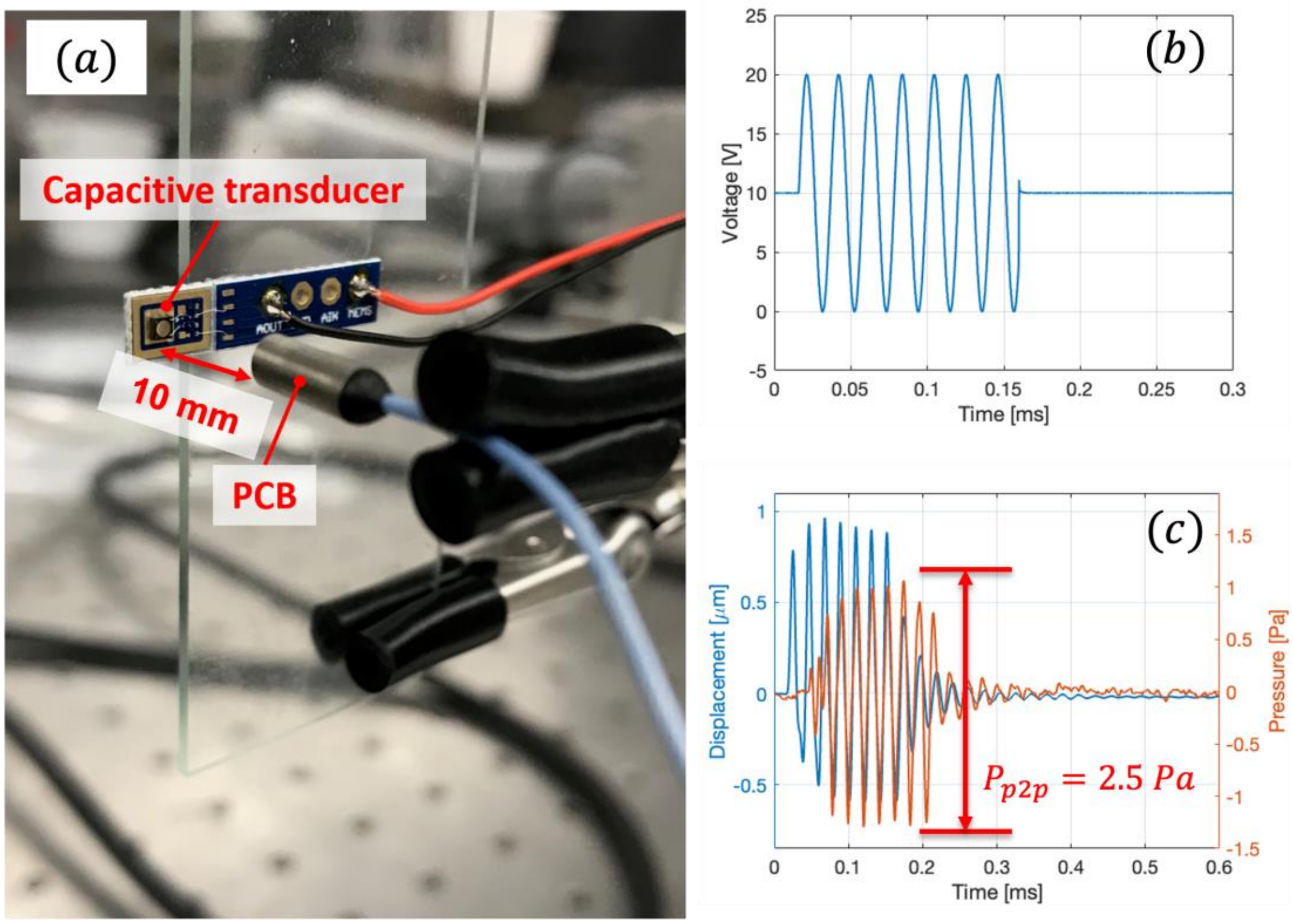

***Figure 4(a) Acoustic waveform measurement setup. (b) A tone burst actuation voltage signal vs. time. (c) Central diaphragm displacement vs. time (blue) and sound pressure measured by the PCB microphone vs. time (orange).***

To demonstrate acoustic waveforms in air, we measured pressure waveforms using a 1/8-inch diameter Piezotronics microphone model PCB 132B38 in combination with a unity gain amplifier model PCB PIEZOTRONICS 482C SERIES. The microphone has a frequency response spanning 11 $kHz$ to 1 $MHz$ and calibrated sensitivity equal to 15.7 $\mu V/Pa$. At 48 $kHz$, the free-field sensitivity of the microphone is approximately 2× the calibrated absolute pressure sensitivity, owing to the reflection of incident sound from the microphone's face.[17,25] A second stage amplifier model Stanford Research Systems SR560 with gain setting of 1,000× (or 60 $dB$) was also used. Considering the diffraction correction and the gain, the sensitivity is taken as 31.4 $\mu V/Pa$. The PCB microphone was placed 10 $mm$ from the MEMS die under study, as shown in Figure 4(a). Figure 4(b) presents the actuation signal, which is a six-cycle sinusoidal tone burst at 48 $kHz$ with voltage ranging from 0 to 20 Volts. The resulting diaphragm central displacement waveform is shown by the blue trace in Figure 4(c). The measured voltage waveform from the calibrated microphone was converted to a pressure waveform using the known sensitivity and is shown by the orange trace in Figure 4(c). The short time delay observed in the pressure signal with respect to the excitation signal is consistent with the expected delay due to the time of flight across a 10 $mm$ distance. The observed peak-to-peak sound pressure is 2.23 $Pa$, or SPL = 92 $dB$.

Relations from physical acoustics can be used to predict on-axis sound pressure resulting from motion of the transducer's surface. For a planar source in an infinite right baffle, the on-axis pressure magnitude $|P|$ in the far field (i.e., past Rayleigh distance) is given by

$$|P| = \frac{\rho}{2\pi r}\omega^2\eta \tag{7}$$

where $\rho$ is the density of air, $r$ is distance from the source, $\omega$ is the angular frequency, and $\eta$ is volumetric displacement amplitude generated by the source. References[26–28] give the general far-field expression for a baffled planar source. Using Eq. 7, the predicted pressure is 2.04 $Pa$, which agrees reasonably well with the measured value.

## 5. PARAMETRIC ARRAY PROTOTYPE

For the parametric-array prototype in air, we used another MEMS microphone with a square diaphragm area. Figure 5(a) shows a top-view photograph of such a MEMS die. The die has a side length of 1.2 $mm$ and contains a $0.62 \times 0.62\ mm^2$ diaphragm. The structure is identical to that in Figure 1(b), except for a 3.4 $\mu m$ gap distance and 20-Volt pull-in voltage. Figure 5(b) shows measured center-point diaphragm-displacement transfer function. This device has a resonance near 88 $kHz$ with half-power bandwidth spanning from 70 to 110 $kHz$.

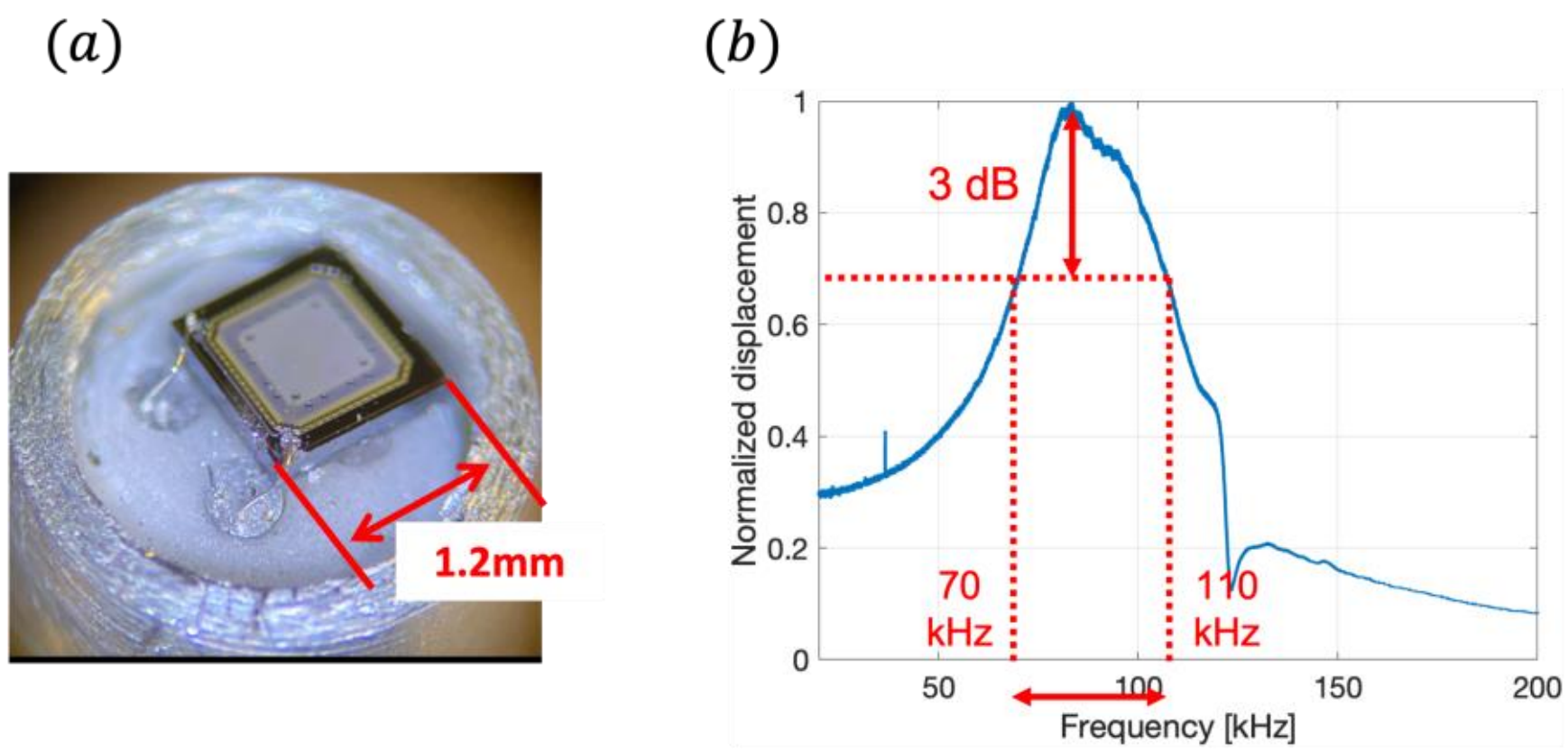

***Figure 5(a) A photograph taken from the top side of a MEMS microphone die with square diaphragm. (b) Measured center-point diaphragm-displacement transfer function driven in small-signal mode.***

A parametric array prototype was built by assembling 28 dies on a printed circuit board (PCB) shown in Figure 6(a). The total array footprint is $9 \times 14\ mm^2$. Because we drive all MEMS dies in pull-in mode, the diaphragms exhibit strongly nonlinear vibration, which dominates the nonlinearity of sound propagation in air. Transducer nonlinearity generates difference frequency omnidirectionally; propagation nonlinearity generates difference frequency directionally. Thus, to get directional low-frequency sound, we have to drive different dies at different primary frequencies. This is common for electrostatic devices. Wygant[22] drove separate wafer quadrants with separate primary signals to prevent nonlinearities from directly radiating sound at the difference frequency. For our device, the 1$^{st}$ and 3$^{rd}$ rows generate one primary frequency; the 2$^{nd}$ and 4$^{th}$ rows generate the other primary frequency. We drove the two groups of dies separately with sinusoidal waveforms ranging from 0 to 24 Volts. All dies operate in pull-in mode. Acoustic pressure was measured at 8.5 $mm$ distance from the device. We first performed this measurement while driving only the 1$^{st}$ and 3$^{rd}$ rows at different frequencies. We then repeated the measurement with only driving the 2$^{nd}$ and 4$^{th}$ rows. Figure 6(b) presents results of the two measurements as the blue and orange curves, respectively. It is worth noting that acoustic pressure response shown in Figure 6(b) seems different from displacement response shown in Figure 5(b). Equation 7 predicts acoustic pressure is proportional to displacement. We attribute this apparent discrepancy to two main factors. (1) Figure 5(b) presents the characteristic of a single MEMS die. Each point in Figure 6(b) plot represents averaging effect across two rows, i.e., 14 dies. The 14 dies may exhibit slight device-to-device variations. (2) Differences among the dies arise partly from manual assembly on the PCB. Since different amounts of epoxy is applied between each die and PCB, the mechanical constraint and intrinsic stress vary from die to die, resulting in the inconsistency between pressure response and diaphragm displacement response. Modern manufacturing is able to control such error into a fairly small value. For example, difference of mechanical sensitivity of each MEMS microphone can be $\pm$ 1 $dBV$. Therefore, uniformity is not expected to be an issue for industrial applications.

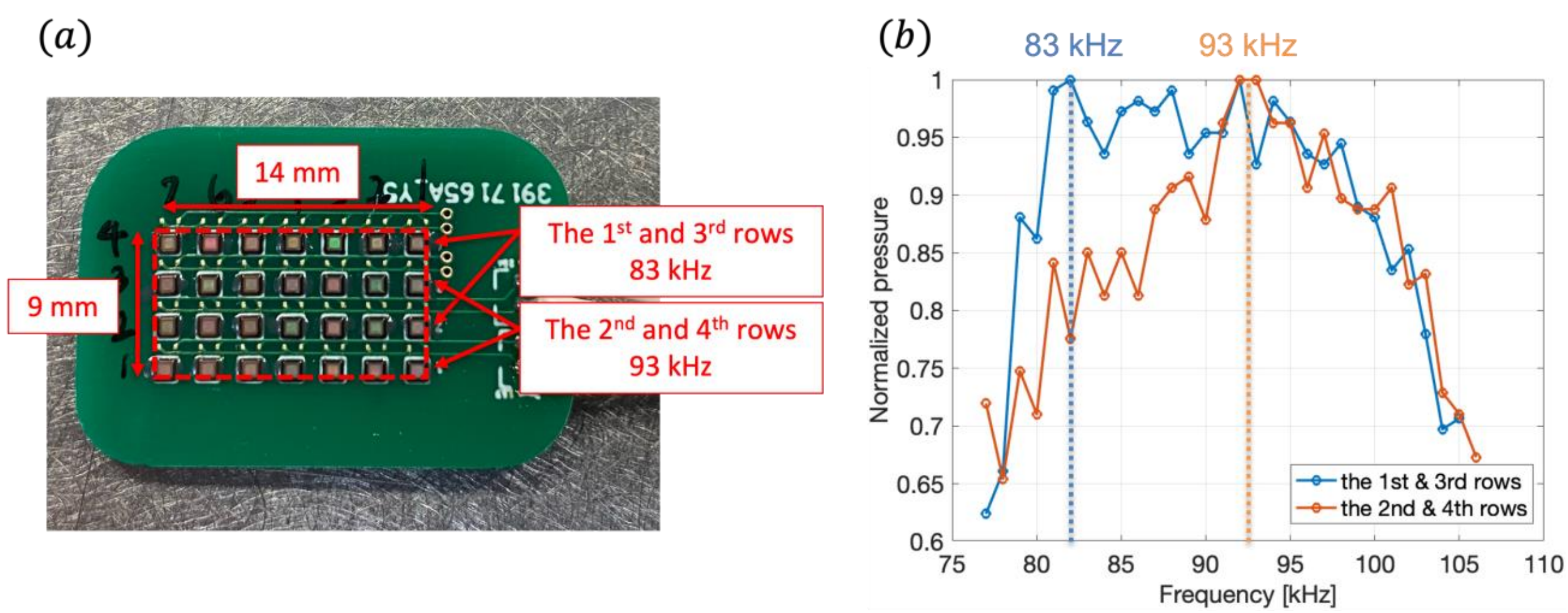

***Figure 6(a) A 28-unit array of the MEMS die built on a PCB. (b) Normalized pressure response for the 1st & 3rd rows (blue plot) and the 2nd & 4th rows (orange plot).***

According to the measured pressure response shown in Figure 6(b), we selected the loudest ultrasonic frequencies as our primary frequencies. Specifically, we actuated the 1st & 3rd rows and 2nd & 4th rows at 83 and 93 $kHz$, respectively. Two signal generators and two power amplifiers were used to drive the PA prototype. Figure 7(a) shows acoustic pressure measurement in the time domain using a PCB microphone and a Polytec MSA-50 data-acquisition system. The resulting waveform shows interference between the two primary frequencies. However, the presence of nonlinearity could be determined only from the frequency-domain spectrum in Figure 7(b). 83 and 93 $kHz$ peaks appear as expected. Note that a peak also appears at 10 $kHz$, which indicates there is nonlinearity as acoustic wave propagates in air. We attribute 10 $kHz$ to propagation nonlinearity rather than another source. We used PCB microphone for this measurement. Because of the high linearity of piezoelectric transduction, receiver-induced nonlinearity is expected to be negligible in the measurement microphone. Nonlinearity could happen in an electrostatic measurement microphone.[20,22] Unfortunately, we cannot characterize the difference frequency quantitatively because bandwidth of the PCB microphone is from 11 $kHz$ to 1$MHz$. Thus, we quantitatively measured primary frequencies and difference frequency using the PCB microphone and G.R.A.S microphone respectively.

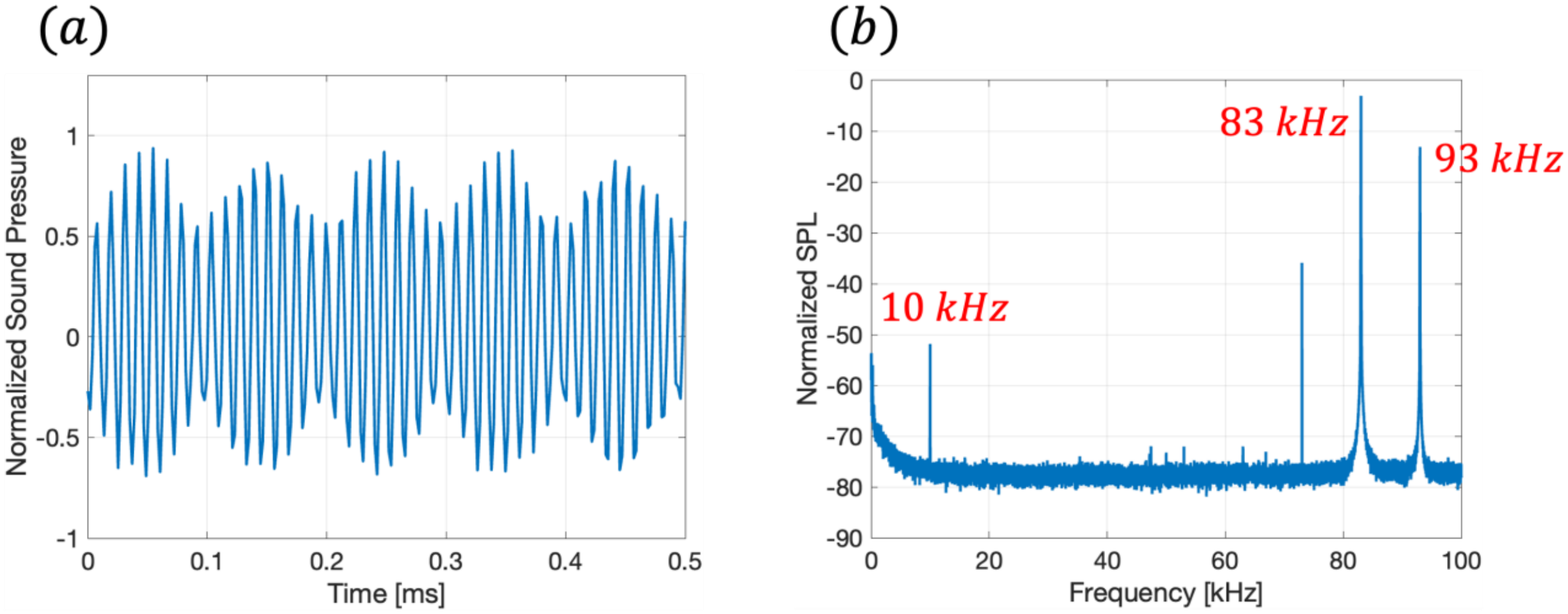

***Figure 7(a) Normalized measured sound pressure waveform in time domain. (b) Normalized measured sound pressure waveform in frequency domain.***

Figure 8(a) shows the measurement setup of difference frequency. The prototype was anchored on a rotation stage secured by a screw. The rotation stage was mounted on a manual stage which enables the prototype to move back and forth. The setup enables the prototype to move with two degrees of freedom. On the right side, the measurement microphone is mounted at the same height as the prototype. A small piece of paper towel was wrapped around the measurement microphone to avoid electrical interference. Note that Figure 8(a) only shows the setup for measuring difference frequency using a G.R.A.S. microphone. We mounted the PCB microphone in the same way for measuring primary frequencies. Figure 8(b) presents pressure angular distribution of both primary and difference frequencies. The N-element array factor[29] in Eq. (8) predicts the half-power beamwidth (HPBW) of the primary frequencies.

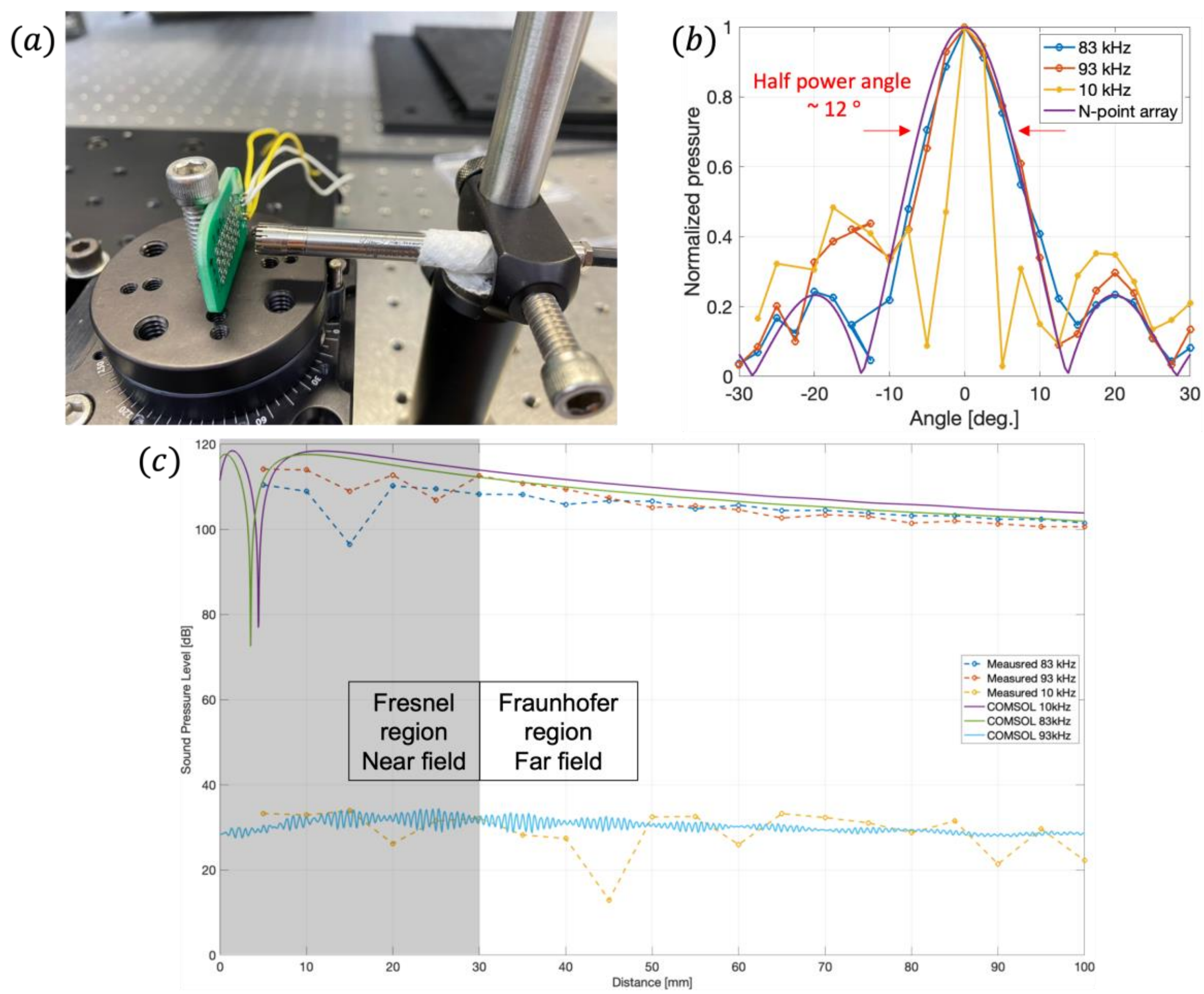

***Figure 8(a) Setup of acoustic pressure measurement. (b) Angular pressure distribution for both measured and modelling waveform. (c) Sound Pressure Level (SPL) measurement and modelling in response to on-axis distance.***

$$\boldsymbol{D(\phi)=\frac{\sin(N\phi)}{N sin\phi}, \phi=\frac{kd}{2} sin\theta} \tag{8}$$

where $\theta$ is the angle with respect to the perpendicular line to the center of the device. The device faces zero angle (i.e., $\theta = 0°$), as shown in Figure 8(a). The label on the x-axis of Figure 8(b) also represents $\theta$. $k$ is the wavenumber. $d$ is spacing between two dies. $N$ is the number of elements. HPBW of difference frequency (yellow plot) has an angular width comparable to those of the primary frequencies. This indicates that the array generated directional low-frequency sound via parametric array method. 10 $kHz$ tone is audible. During measurement, we could hear the tone when our ear was perpendicular to the PCB. But we heard nothing when the device was turned in other directions.

Figure 8(c) presents sound pressure level (SPL) vs. distance for both primary and difference frequencies. Dashed lines with markers represent measured results. Solid lines represent finite-element-analysis (FEA) results. The top two dashed lines are measured results for primary frequencies. The solid and dashed lines at the bottom are modeling and measurement of the difference frequency. Results show good agreement between the model and measurements. We modeled the PA using the Westervelt equation via FEA. Westervelt equation is valid in far field (i.e., Fraunhofer region or beyond Rayleigh distance) labeled in Figure 8(c).[30]

The yellow dashed line at the bottom is the measured value reduced by 10 $dB$. The 10 $dB$ correction accounts for spurious sound attributed to receiver nonlinearity. The spurious-sound level was not independently determined. Wygant[22] estimated a 10 $dB$ contribution. After applying the 10 $dB$ correction, the model (blue solid curve) and measurements (yellow dashed curve) agree reasonably well.

# 6. DISCUSSION

## A. FILL FACTOR AND DEVICE SIZE LIMITS

Although the prototype demonstrates the PA effect, $SPL$ is only about 35 $dB$, as shown in Figure 8(c). For comparison, $SPL$ of normal conversation is about 60 $dB$. To explore reasons for the inefficiency, we could further expand equation 7.

$$|P| = \frac{\rho}{2\pi r}\omega^2\eta = \frac{\rho}{2\pi r}\omega^2 S_{PA} \cdot FF \cdot u_d \tag{9}$$

where $u_d$ is area-averaged diaphragm displacement, equal to $u_c\alpha$ shown in equation 9. Note that Eq. (9) cannot predict difference-frequency pressure, because it is based on the linear Rayleigh integral. However, such an equation can still aid qualitative understanding. First, $S_{PA}$ is too small. The prototype is $9 \times 14 = 126\ mm^2$; typical area of a laptop speaker is around $4{,}000\ mm^2$. Second, $FF$ is too small. $FF$ is equal to diaphragm area divided by device area. i.e., $FF = \frac{0.62^2 \times 14}{9 \times 14} = 3.25\%$. Square diaphragm side length is 0.62 $mm$. Each primary-frequency tone is generated by 14 elements. Thus, both $S_{PA}$ and $FF$ limit the difference-frequency SPL.

## B. SPURIOUS SOUND EFFECT

Nonlinearity of measurement microphone affects difference frequency measurements[20], an effect referred to as spurious sound[22]. Wygant estimated that 10 $dB$ is a good estimate.[22] After subtracting 10 $dB$, we obtain the blue solid line shown in Figure 8(c).

## C. FEA MODEL

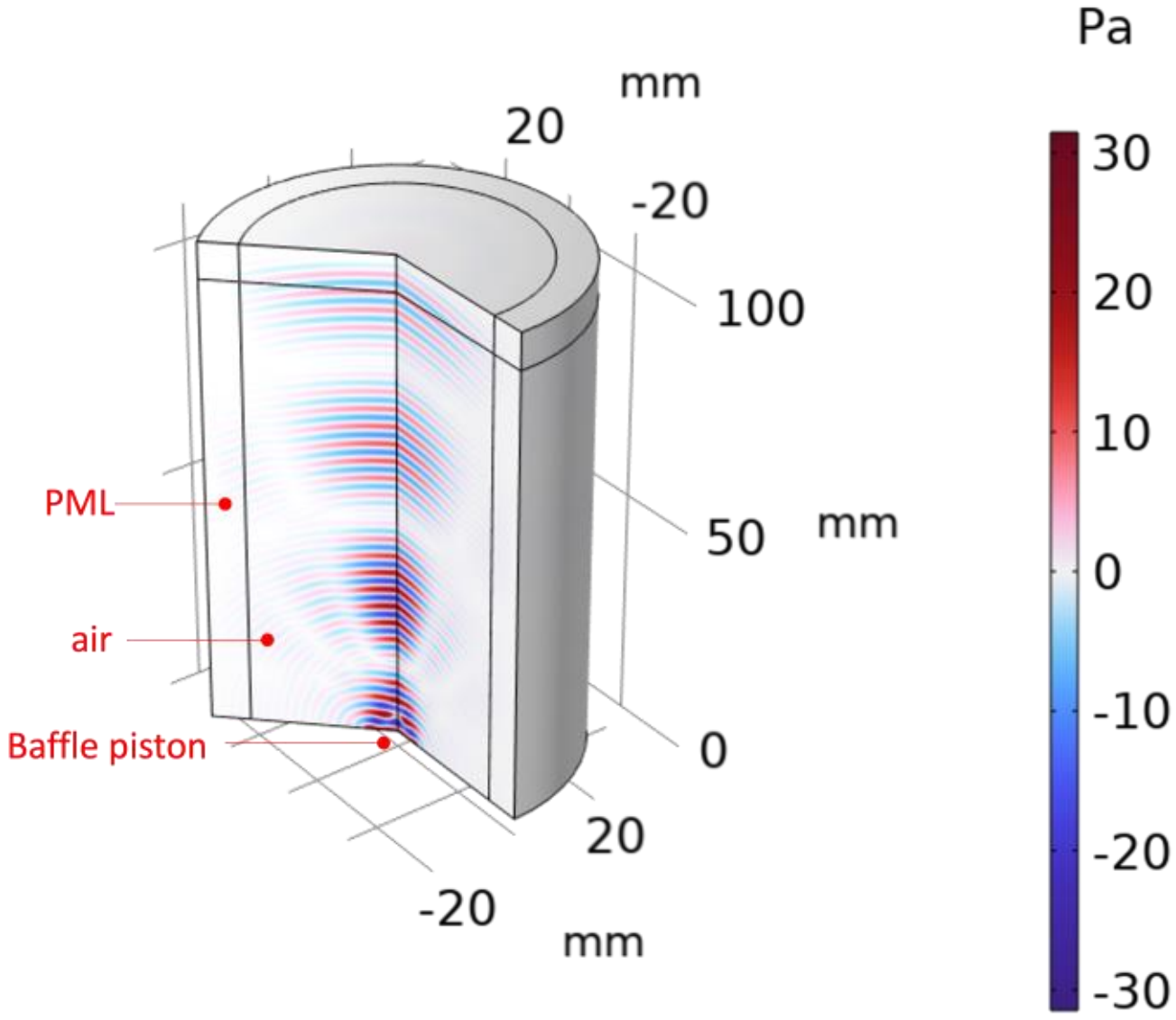

***Figure 9. Spatial acoustic pressure distribution in COMSOL model.***

To model PA, we performed the FEA in COMSOL. We built an axisymmetric model in the time domain using nonlinear-acoustics settings. Figure 9 presents the entire computational domain. A baffled circular piston is at the bottom. The circular piston's radius ($\alpha$) is given by $\pi a^2 = 9 \times 14\ mm^2$. Piston surface displacement is given by equation 9, which gives equivalent volume displacement for both the piston and the PA prototype.

$$S_{pstn} \cdot u_{pstn} = (S_{PA} \cdot FF) \cdot (u_c \cdot \alpha) \quad (10)$$

where $S_{pstn}, u_{pstn}, S_{PA}, FF, u_c$, and $\alpha$ represent the area of piston, displacement of piston, area of the PA prototype, fill factor (ratio of diaphragm area to whole device area), central diaphragm displacement, and effective coefficient (ratio of the area-averaged diaphragm displacement to central diaphragm displacement). The propagation medium is air, as labeled in Figure 9. A perfectly matched layer (PML) surrounds the air domain to reduce reflections. At a representative time, the acoustic pressure in space is shown in the cylindrical region.

### D. DESIGN-SPACE EXPLORATION

Currently, the PA design space remains underexplored because of the limited ultrasonic transducers available on the market. Pull-in CMUTs provide transducers for PA applications over a broader range of primary frequencies, surface velocities, and surface areas than currently available transducers. We will explore the PA design space with respect to primary frequency, surface velocity, and surface area. Modeling the PA in the time domain with COMSOL is computationally expensive. Lee worked on time-domain solution of KZK equation in 1995.[31] In 2019, Cervenka provided a frequency-domain modeling method.[32] Zhong studied spherical wave expansion (SWE) in 2020.[33] These methods reduce the computational cost of nonlinear-acoustics simulations. They also provide a foundation for exploring PA design space.

## 7. CONCLUSION

This work demonstrates a pull-in actuation mode for CMUTs by using a MEMS microphone structure. Pull-in mode enables the diaphragm to traverse the full gap between the diaphragm and the backplate. Large diaphragm displacement generates large acoustic pressure in air. Using such large acoustic pressure, the parametric-array effect was demonstrated. Modeling and measurements are in good agreement. With further development of pull-in CMUTs, a PA directional speaker could potentially be integrated into consumer electronics.